\documentclass[apl,twocolumn,superscriptaddress]{revtex4-1}
\usepackage[latin1]{inputenc}
\usepackage{amsmath}
\usepackage{amsfonts}
\usepackage{amssymb}
\usepackage{graphicx}
\usepackage{float} 

\begin{document}

\title{Ab initio calculation of Spin-Polarized Low-Energy Electron Diffraction Pattern\\ for
the systems Fe(001) and Fe(001)-p(1x1)-O}
\date{\today}
\author{Stephan Borek}
\author{J\"urgen Braun}
\affiliation{Department Chemie, Ludwig-Maximilians-Universit\"{a}t M\"{u}nchen, Butenandtstra\ss e 5-13, 81377 M\"{u}nchen, Germany}
\author{J\'an Min\'ar}
\affiliation{Department Chemie, Ludwig-Maximilians-Universit\"{a}t M\"{u}nchen, Butenandtstra\ss e 5-13, 81377 M\"{u}nchen, Germany}
\affiliation{New Technologies-Research Centre, University of West Bohemia,\\ Univerzitni 8, 306 14 Pilsen, Czech Republic}
\author{Hubert Ebert}
\affiliation{Department Chemie, Ludwig-Maximilians-Universit\"{a}t M\"{u}nchen, Butenandtstra\ss e 5-13, 81377 M\"{u}nchen, Germany}

\begin{abstract}
The construction of a multi-channel vector spin polarimeter requires the development of a detector type, which works as a
spin polarizing mirror with high reflectivity and asymmetry properties to guarantee for a high figure of merit. Technical
realizations are found by spin polarized electron scattering from a surface at low energies. A very promising candidate
for such a detector suitable material consists of an oxygen passivated iron surface, as for example a Fe(001)-p(1x1)-O surface. We investigate in 
detail the electronic structure of this adsorbate system and calculate the corresponding spin-polarized low-energy electron
scattering. Our theoretical study is based on the fully relativistic SPRKKR-method in the framework of density functional theory.
Furthermore, we use the local spin-density approximation in combination with dynamical mean field theory to determine the
electronic structure of Fe(001)-p(1x1)-O and demonstrate that a significant impact of correlation effects occurs in the
calculated figure of merit.
\end{abstract}

\maketitle

\section{Introduction}

The first quantitative theoretical description of relativistic spin-polarized low-energy electron diffraction (SPLEED)
had been developed by Feder (1983) and by Tamura and Feder (1984) \cite{feder3,tamura1}. The advantage of a
spin-polarized relativistic formulation consists in the fact that the interplay of exchange interaction and spin-orbit coupling
is considered on the same level of accuracy \cite{celotta1,feder1}. The application of this method to electron scattering
from solid surfaces allows to support the development of a multi-channel vector spin polarimeter which can be realized 
using selected surfaces as two-dimensional reflection mirrors. The concept of such a detection method using spin-dependent
electron scattering has been demonstrated, for example, for W(100) \cite{kolbe1}. While in the cited work the detection
of only one spin component has been realized the new scattering mirror should give the possibility to detect all three
spin components in a single step. Therefore, an optimization of suitable materials for the use as reflection mirrors and
corresponding investigations of new materials is highly desirable. The actual research activities focus on two classes
of single crystal surfaces: non-magnetic surfaces from high $Z$-materials, where spin-orbit coupling acts as the underlying
physical mechanism or magnetic surfaces (ferromagnetic materials) where both exchange interaction and spin-orbit coupling
influence the spin-dependent electron scattering. The classical system representing the spin-orbit case is W(100)
\cite{kirschner1}. Another promising candidate was found in the Ir(001) surface, which is less reactive than tungsten and,
as a consequence, provides a longer operation time for the use as a spin detector \cite{kutnyakhov1}. In contrast the scattering
at ferromagnetic surfaces has been investigated only recently \cite{hillebrecht1}. Nevertheless, it was shown that using
ferromagnetic materials as spin-detectors a very high figure of merit (FOM) can be achieved \cite{bertacco2}.

In the last decades significant theoretical and experimental progress was made in the application of such systems. One
major success of these investigations was the determination of surface magnetic moments \cite{feder1}. For the exchange
scattering of electrons from a sample surface different ferromagnetic materials have been used
\cite{hillebrecht1,bertacco1,winkelmann1,okuda1}. An often mentioned problem of spin detectors which depend on spin-orbit
interaction is the low FOM in the order of $10^{-4}$ \cite{kutnyakhov1}. A higher FOM was reported for exchange scattering
from an iron surface where values up to 20 times larger have been reached \cite{hillebrecht1,bertacco1,winkelmann1,okuda1}.
A disadvantage of these surfaces is their short operation time due to contamination. A solution to this problem is the
preparation of an oxygen overlayer, i.e. a surface passivation. For the coverage of 1 monolayer oxygen an ordered overlayer
is formed resulting in p(1x1) LEED reflection patterns, leading therefore to a longer operation time in vacuum \cite{bertacco1}.
In this theoretical study the Fe(001)-p(1x1)-O surface serves as a benchmark for further developments of our theoretical
approach, as well as an suitable starting point for research activities on various materials which may applicable as
reflection mirrors for spin filtering.

The paper is organized as follows: In Sec. \ref{sec_theory} we describe the theory of our SPLEED calculations.
In Sec. \ref{sec_discussion} we discuss our theoretical method concerning the electronic structure and the various SPLEED
calculations and in Sec. \ref{sec_summary} we summarize our results.

\section{Theory \label{sec_theory}}
\subsection{Electronic structure}
The calculation of the electronic structure has been done using the Munich SPRKKR program package \cite{korringa1,kohn1,ebert1,SPR-KKR6.3}.
The implementation of the tight-binding (TB) KKR-method allows an effective treatment
of two-dimensional surfaces,
i.e. the self-consistent calculation of the electronic structure,
due to the fast convergence of the TB structure constants \cite{korringa1,kohn1,kambe1,ebert2,lovatt1,feder1}. These
decay exponentially which allows in particular the treatment of various layered systems and 
relaxed surfaces with adsorbed atoms \cite{zeller1}. 
Using this method we construct a semi-infinite system with two-dimensional periodicity
which consists of three parts: substrate (having bulk potential), surface region
and vacuum region (represented by empty spheres).
The calculations were done fully relativistic to treat effects coming from spin-orbit coupling
and exchange interaction in a coherent way.
To account for many body effects beyond the local spin density approximation \cite{HK64,KS65,SK66} 
a site diagonal, nonlocal, complex and energy dependent self energy $\Sigma$
determined within the dynamical mean field approach (DMFT) \cite{KV04,G96} has been used.
It has been shown that this method is straightforwardly
applicable to semi-infinite lattices with lateral translational invariance and
an arbitrary number of atoms per unit cell \cite{minar1,braun1}.
The inclusion of many-body effects expressed by the DMFT
has been shown to result in significant changes for the
shape of the calculated spectra especially for lower kinetic energy
of the reflected electrons \cite{braun1}.
An important parameter for various spectroscopic calculations is the work function.
We calculated the work function of the 2D semi-infinity system applying
a summation over the Madelung potentials in the interaction zone as described
elsewhere \cite{zabloudil1}. 

\subsection{Theory of SPLEED \label{theory_spleed}}
The different asymmetries that characterize a SPLEED spectrum of a ferromagnetic surface 
are determined by changing the magnetization direction parallel to
the surface either in the scattering plane or perpendicular to it. The scattering plane is defined by the
wave vector of the incident and the scattered electrons. Beside the change of the magnetization
the polarization of the incident electrons has to be changed resulting in four different scattered intensities \cite{feder1,tamura1}. 
These are determined by the electron polarization ($\sigma$) as well as the direction of the magnetization ($\mu$).
It has been shown that mainly two different set-ups have to be considered concerning the orientation of magnetization and polarization 
with respect to the scattering plane \cite{tamura1}.
For our calculations we used the setup for which both magnetization and polarization are parallel to the scattering plane.
If this plane is parallel to a mirror plane the spin-orbit asymmetry vanishes and the scattering of the electrons is only due
to exchange interaction \cite{feder1}. According to symmetry considerations for the
scattered electron intensities 

\begin{equation}
 I_{\mu}^{\sigma}=I_{-\mu}^{-\sigma} \label{eq:symmetry_intensity}
\end{equation}

holds \cite{tamura1}. For the spin-orbit asymmetry ($A_{soc}$) defined by \cite{feder1}

\begin{equation}
 A_{soc}=\frac{1}{2}(A_{+}-A{-}).
\end{equation}

with the definition for $A_{+}$ and $A_{-}$

\begin{align}
 A_{+}=\frac{I_{+}^{+}-I_{+}^{-}}{I_{+}^{+}+I_{+}^{-}} \\
 A_{-}=\frac{I_{-}^{+}-I_{-}^{-}}{I_{-}^{+}+I_{-}^{-}}.
\end{align}

$A_{soc}=0$ results. The exchange asymmetry ($A_{ex}$) in turn can be expressed by \cite{feder1}

\begin{equation}
 A_{ex}=\frac{1}{2}(A_{+}+A_{-}). \label{eq:exchange_asymmetry}
\end{equation}

Based on the symmetry restriction in Eq. (\ref{eq:symmetry_intensity}) one can evaluate
$A_{ex}$ from the following simplified equation

\begin{equation}
 A_{ex}=\frac{I_{+}^{+}-I_{+}^{-}}{I_{+}^{+}+I_{+}^{-}}.
\end{equation}

As a consequence, the scattering plane is parallel to a mirror plane only one magnetization direction
has to be considered when determining the exchange asymmetry. Nevertheless a useful test is to consider
in addition the reversed magnetization to verify vanishing spin-orbit asymmetry. 
%
%

Another quantity to characterize different working points or regimes for surfaces
used as scattering mirror is the FOM. It is defined as the
product of the reflected intensity and the asymmetry for a specific orientation of the
magnetization:

\begin{equation}
	\text{FOM}_{+(-)}=I_{+(-)}\cdot A_{+(-)}^{2}. \label{eq:fom}
\end{equation}

Here the indices indicate the magnetization direction of the sample. For the use as a spin-filter
both reflectivity and asymmetry should have high values, leading to a high FOM.
%
%
%
%

\section{Discussion \label{sec_discussion}}
\subsection{Electronic structure calculation}
The calculation of the electronic structure has been started from a fully
relaxed surface and interface using experimental structure parameters \cite{chubb1}.
For surface sensitive methods it is important to include the structural
relaxation of the topmost surface layers. 
Especially for methods using low energetic particles the changes in the electronic
structure resulting from the relaxations are important.
We set up a two-dimensional surface system which consists of 10 monolayers
(ML) Fe, 1 ML O and 9 ML empty spheres to simulate the vacuum.
In terms of the TB-KKR for 2D systems we introduce left
and right bulk regions representing the properties of the
Fe substrate and the vacuum.
For the left bulk region we used 2 ML of Fe repeated to the left while
2 ML of the topmost empty spheres have been used for simulating
the right bulk region, i.e. the vacuum region.
Considering the electronic structure the interaction zone in between simulates
the transition from Fe bulk to surface properties.
Our calculations were done
in the atomic sphere approximation (ASA) using the parameterization for the
exchange-correlation functional of Vosko, Wilk and Nusair \cite{Vosko1}.
We used a lattice constant of 2.86 $\r{A}$ according to the unit cell of bcc Fe.
A fully relativistic ab initio calculation of the potentials was applied to
account for spin-orbit and exchange effects on one footing. 
We also included many body effects in our calculations considering the
sensitivity for spectroscopies based on low energetic particles.
For solving the many-body effective impurity problem a DMFT solver has been used \cite{kotliar1,held1}.
In our calculations we utilized the spin-polarized $T$-matrix approximation solver (TMA)
\cite{chadov1}. It has been shown that the use of a TMA solver
is justified because of the less pronounced correlation effects in transition metals \cite{sipr1}.
The parameters which have to be supplied for a DMFT calculation is the intra-atomic
Hund exchange  interaction ($J$) and the screened Coulomb interaction ($U$).
Corresponding to previous extensive studies we set the values
to $J=0.9$ eV and $U=2.3$ eV \cite{sipr1}.

\begin{figure}[h]
	\includegraphics[scale=0.4,clip=true,trim=0cm 5cm 0cm 4cm]{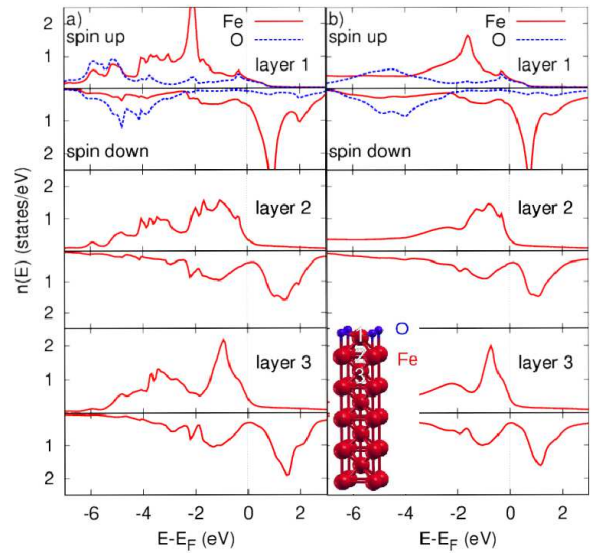}
\caption{DOS of the first three atomic layers of the investigated half-infinity surface system
         Fe(001)-p(1x)-O.
         In a) the DOS for a standard LSDA calculation is shown. In b) the calculation includes
	 many-body effects accounting via LSDA+DMFT.
	 The DOS of O and the first Fe layer are drawn together in the uppermost panel to show the
         hybridization of O and Fe in the valence band. The inset shows a sketch of the surface
         system and the numbering of the atomic layers.}
	 \label{dos}
\end{figure}

The results of the electronic structure calculations are shown in Fig. \ref{dos} in terms
of the density of states (DOS) of the first three Fe layers and the O layer. 
Beside the result of LDA calculation the DOS of a DMFT calculation is shown as well.

The DOS shows reasonable agreement with previous electronic structure calculations for the system
Fe(001)-p(1x1)-O \cite{tange1}.
It is visible that in the energies regime $-6$ eV to $-2$ eV a hybridization
of the valence states between O and the Fe layer occurs.
The LSDA+DMFT based DOS calculations show a broadening especially for the
topmost Fe layers resulting in spectral changes for surface sensitive
spectroscopic methods.
This is caused by the finite value of the self
energy in the specific energy range and is also visible in the
calculated band structure (see Fig. \ref{bsf_dmft} and Fig. \ref{selfenergy_ImSig}).
The changes in both spin channels result in a lower spin magnetic
moment ($m_{spin}$) for the DMFT calculations. 
The main features agree
using LSDA or LSDA+DMFT indicating that the main properties
of the electronic and magnetic structure of the Fe(001)-p(1x1)-O
surface can be described using both schemes.
Nevertheless, for the calculation of very low-energy
electron diffraction it is important to include LSDA+DMFT (see below).
In Fig. \ref{magnetic_moments} the
spin- and orbital magnetic moments for bcc Fe bulk and the three topmost
atomic layers of Fe(001)-p(1x1)-O are shown.
The Fe atoms of the topmost layer have a higher magnetic moment
when compared to the bulk value. This finding is known for
magnetic atoms on surfaces and is related to the band narrowing of the $d$-states \cite{tamura2,fu1}.
The trend for the decrease of the spin magnetic moments
going to deeper Fe layers is similar for LSDA and LSDA+DMFT calculations. 
This behavior is reflected by the comparable main features
in the DOS (see Fig. \ref{dos}) comparing LSDA and LSDA+DMFT.

\begin{figure}[H]
\begin{minipage}[c]{.2\textwidth}
\includegraphics[scale=0.22,angle=270,clip=false,trim=0cm 4.cm 1.cm 7.5cm]{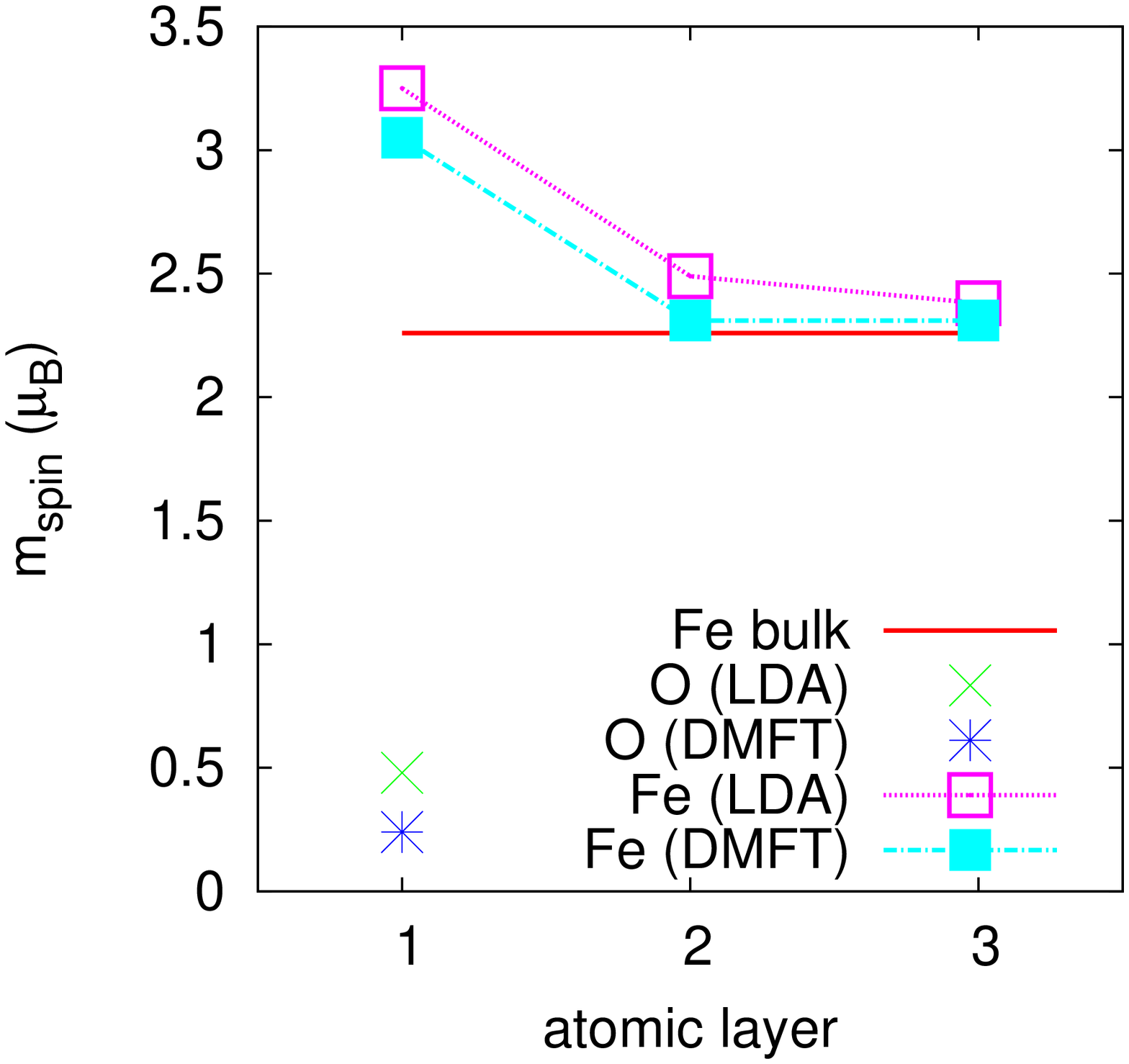}
\end{minipage}
\hspace{1.0cm}
\begin{minipage}[c]{.2\textwidth}
\includegraphics[scale=0.22,angle=270,clip=false,trim=0cm 6.2cm 1.cm 4cm]{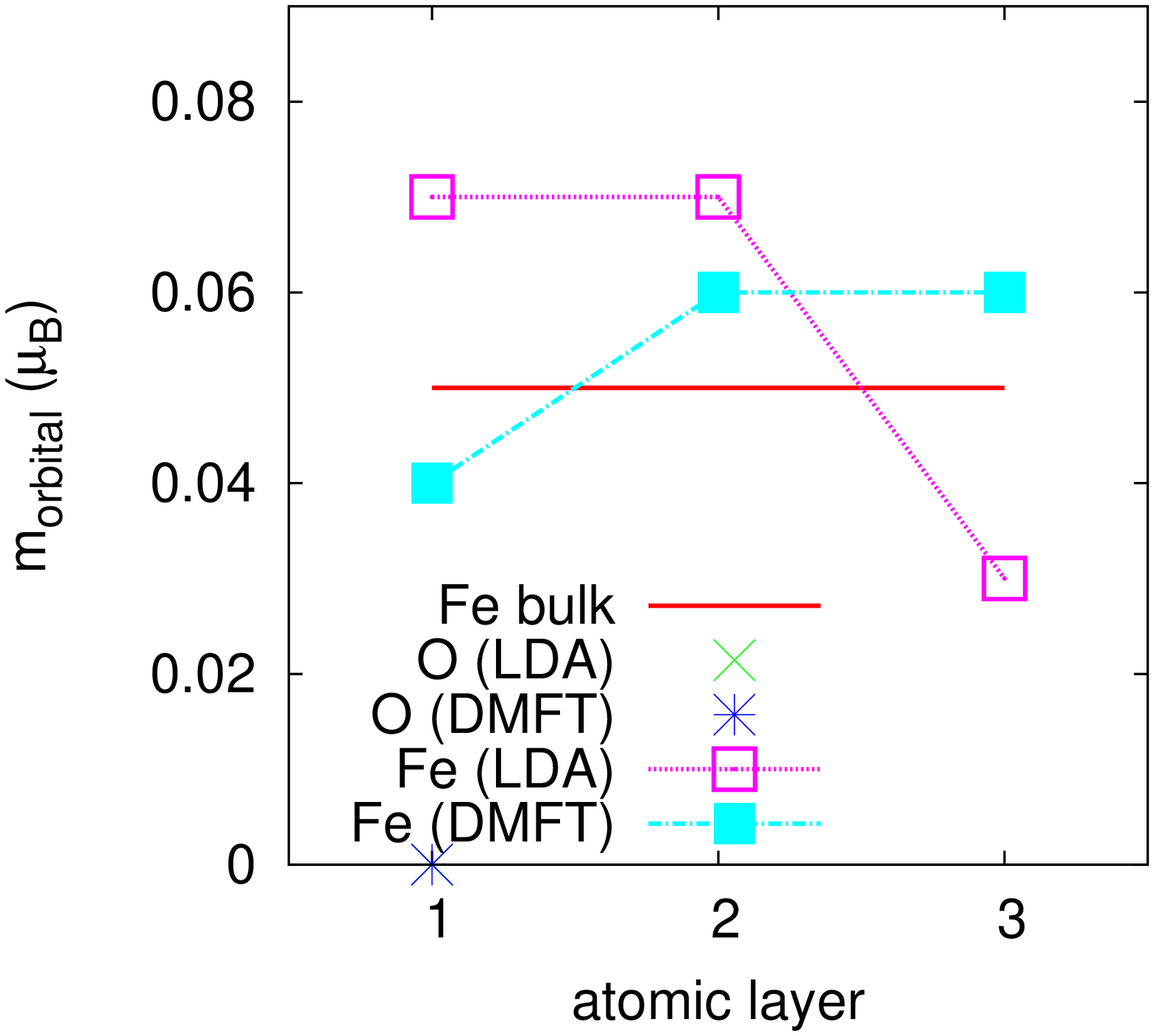}
\end{minipage}
\caption{Magnetic moments for Fe bulk and the first three atomic layers of the semi-infinite system
         Fe(001)-p(1x1)-O with and without the inclusion of many-body effects. Beside the
         spin magnetic moments the orbital moments are shown. For O the orbital moment
         has zero value for LDA and DMFT.}
\label{magnetic_moments}
\end{figure}

Due to the hybridization of O and Fe
in the valence band regime a magnetic moment for O is induced.
The increased magnetic moment of Fe for the surface layer in combination with
the induced magnetic moment for O results in a
larger exchange interaction at the passivated Fe(001) surface
in contrast to a clean Fe(001) surface.
Comparing the spin-orbit induced orbital magnetic moments the differences
between LSDA and LSDA+DMFT calculations are more pronounced than for the spin moments.
Using the LSDA an enhancement of
the orbital moment for the two outermost atomic layers occurs as
one would expect. Beside the decrease of the orbital moment going
to deeper Fe layers is stronger for the LSDA calculation.
Nevertheless the dominating part is the spin magnetic moment which
characterises the exchange scattering of the polarized electrons.
In summary the passivation of the Fe surface results in a significant change of
the magnetic properties compared to a non-passivated Fe surface. 
In our calculations for the first Fe layer of a non-passivated Fe surface a
spin and orbital magnetic moment $m_s=2.81 \mu_B$ and $m_o=0.11 \mu_B$ results.
Whereas for a O passivated surface using the LSDA for the first Fe layer
$m_s=3.25 \mu_B$ and $m_o=0.07 \mu_B$ have been calculated.
The enlarged magnetic moments results in an increased exchange scattering
as been shown in previous experiments \cite{bertacco1}. 
Also a very high Sherman function was reported 
which is linked directly to the magnetic properties of the surface \cite{bertacco2}.
It should be mentioned that our calculated magnetic moments especially
for the two surface layers show good agreement with data in the
literature \cite{tange1}.

\subsection{SPLEED calculation}
\begin{figure}[H]
\begin{minipage}[c]{.2\textwidth}
\includegraphics[scale=0.3,angle=270,clip=true,trim=2cm 4.cm 1.cm 7.5cm]{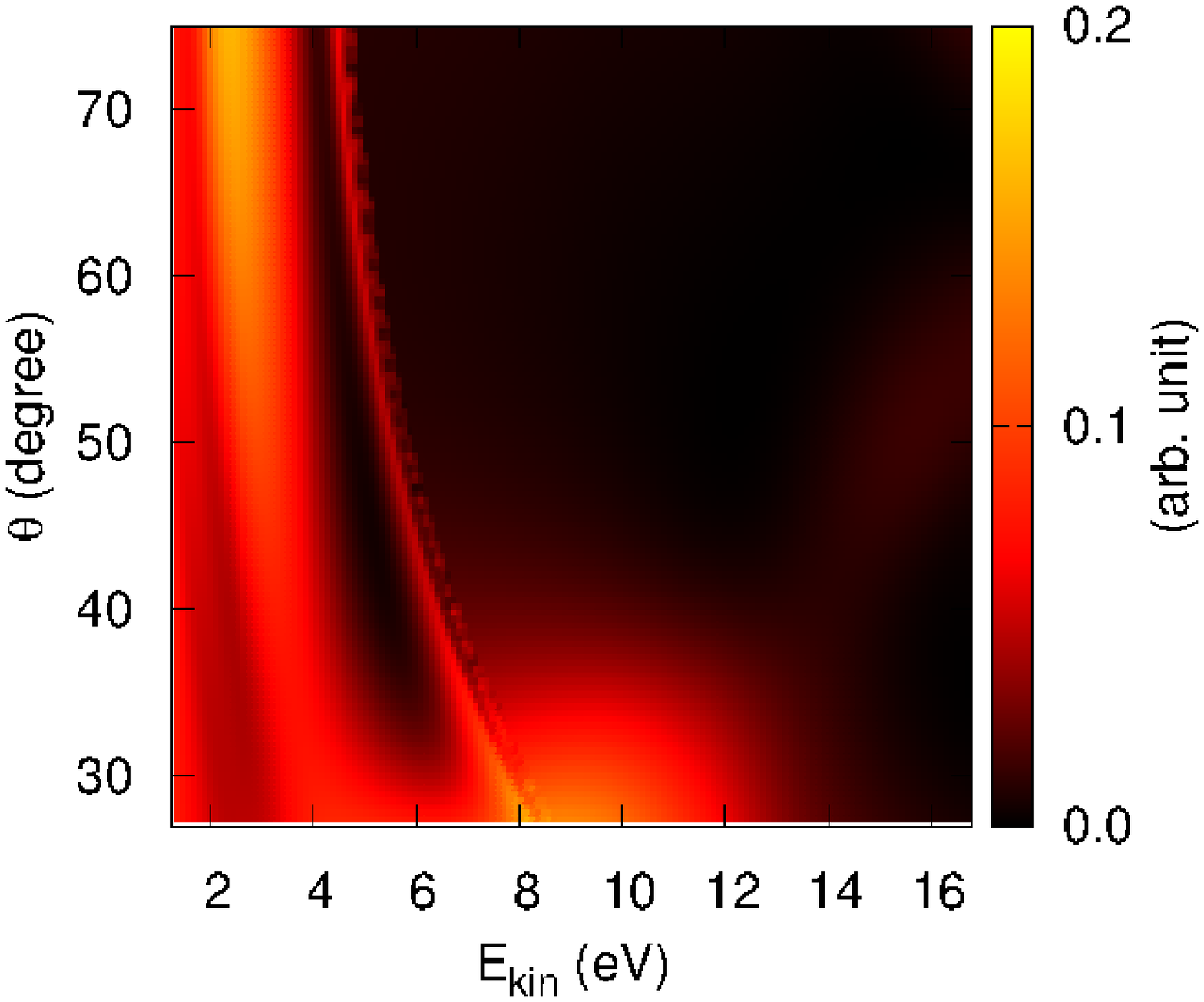}
\end{minipage}
\hspace{0.5cm}
\begin{minipage}[c]{.2\textwidth}
\includegraphics[scale=0.3,angle=270,clip=true,trim=2cm 6.2cm 1.cm 4cm]{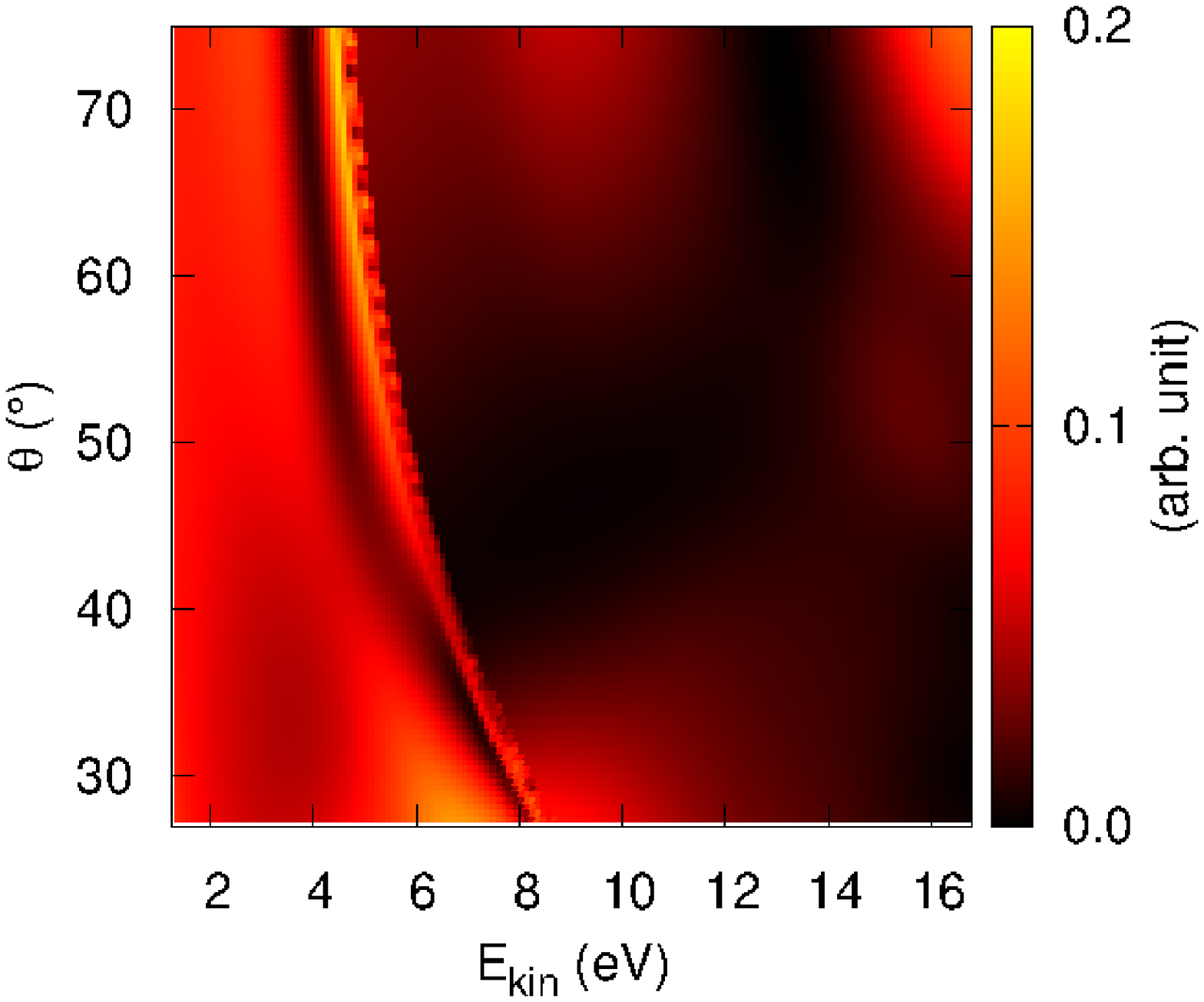}
\end{minipage}
\caption{Left: $\Theta$-energy-map of the reflectivity for Fe(001). Right: The same for Fe(001)-p(1x1)-O.}
\label{theta_energy_maps_refl}
\end{figure}

\begin{figure}[H]
\begin{minipage}[c]{.2\textwidth}
\includegraphics[scale=0.3,angle=270,clip=true,trim=2cm 4.cm 1.cm 7.5cm]{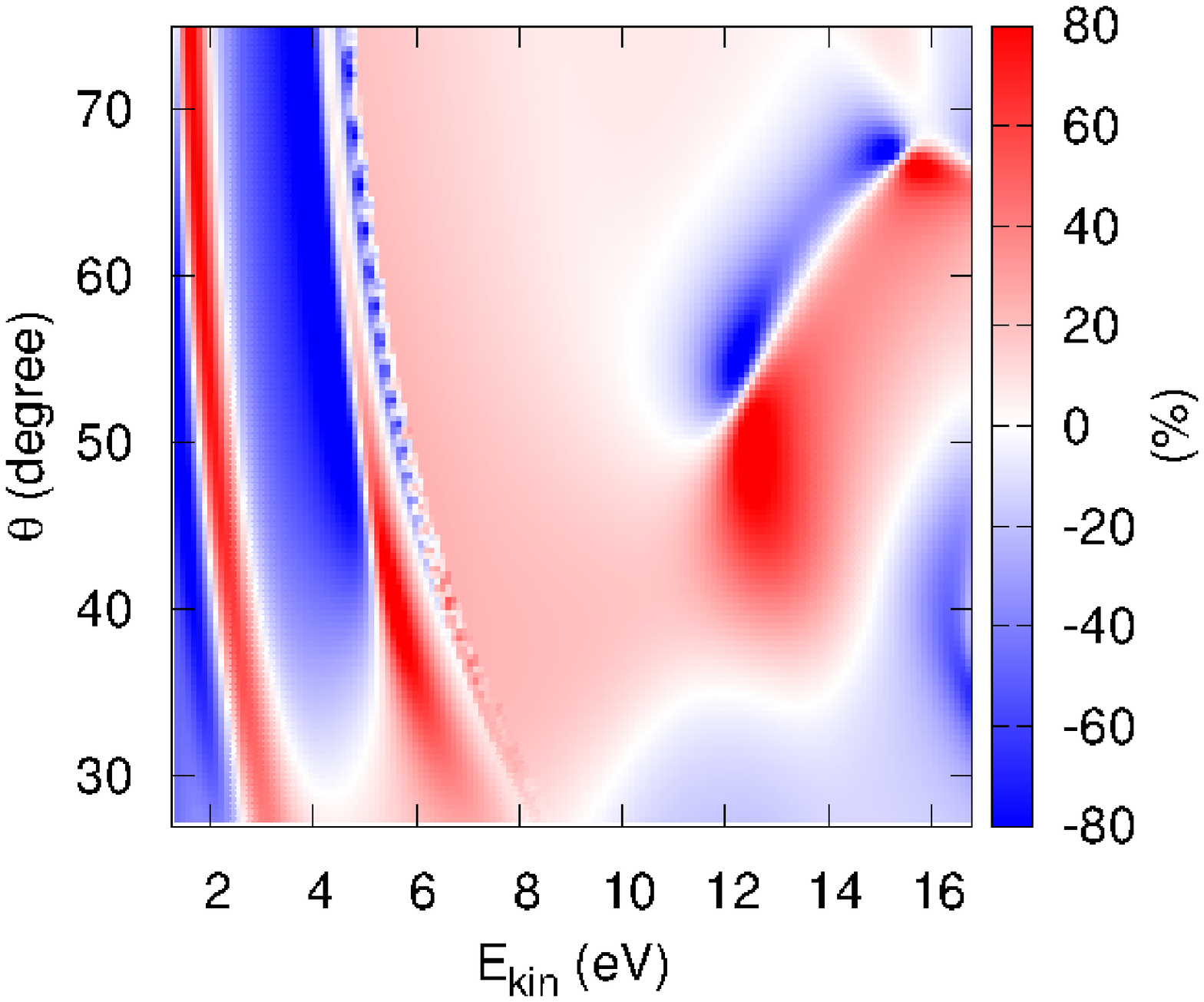}
\end{minipage}
\hspace{0.5cm}
\begin{minipage}[c]{.2\textwidth}
\includegraphics[scale=0.3,angle=270,clip=true,trim=2cm 6.2cm 1.cm 4cm]{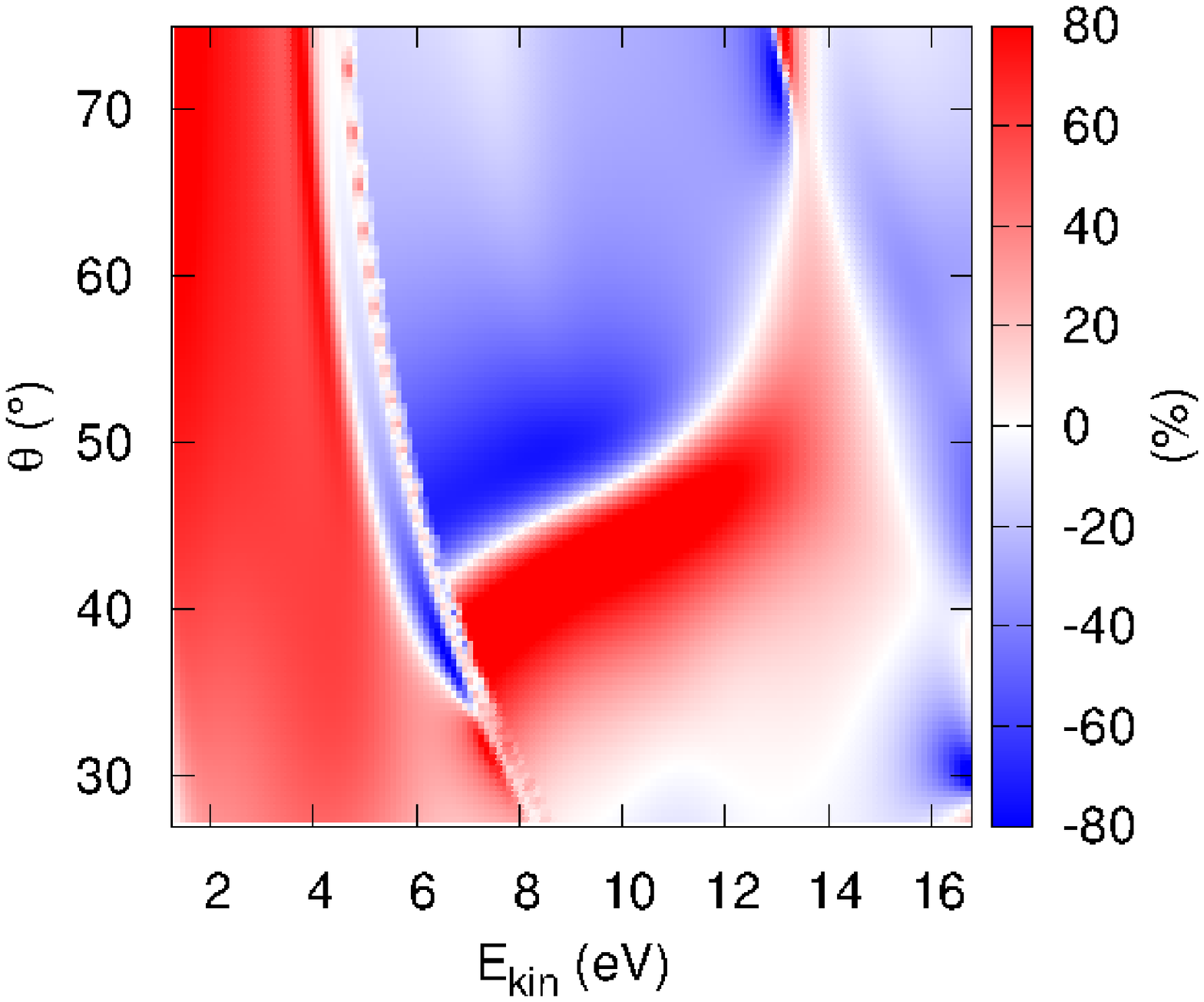}
\end{minipage}
\caption{Left: Exchange asymmetry ($A_{+}$) for Fe(001).
Right: The same for Fe(001)-p(1x1)-O.}
\label{theta_energy_maps_asym_p_m_fe001_fe001_O}
\end{figure}

\begin{figure}[h]
\hspace{-1cm}
\begin{minipage}[c]{.2\textwidth}
\includegraphics[scale=0.3,angle=270,clip=true,trim=2cm 4.cm 1.cm 7.5cm]{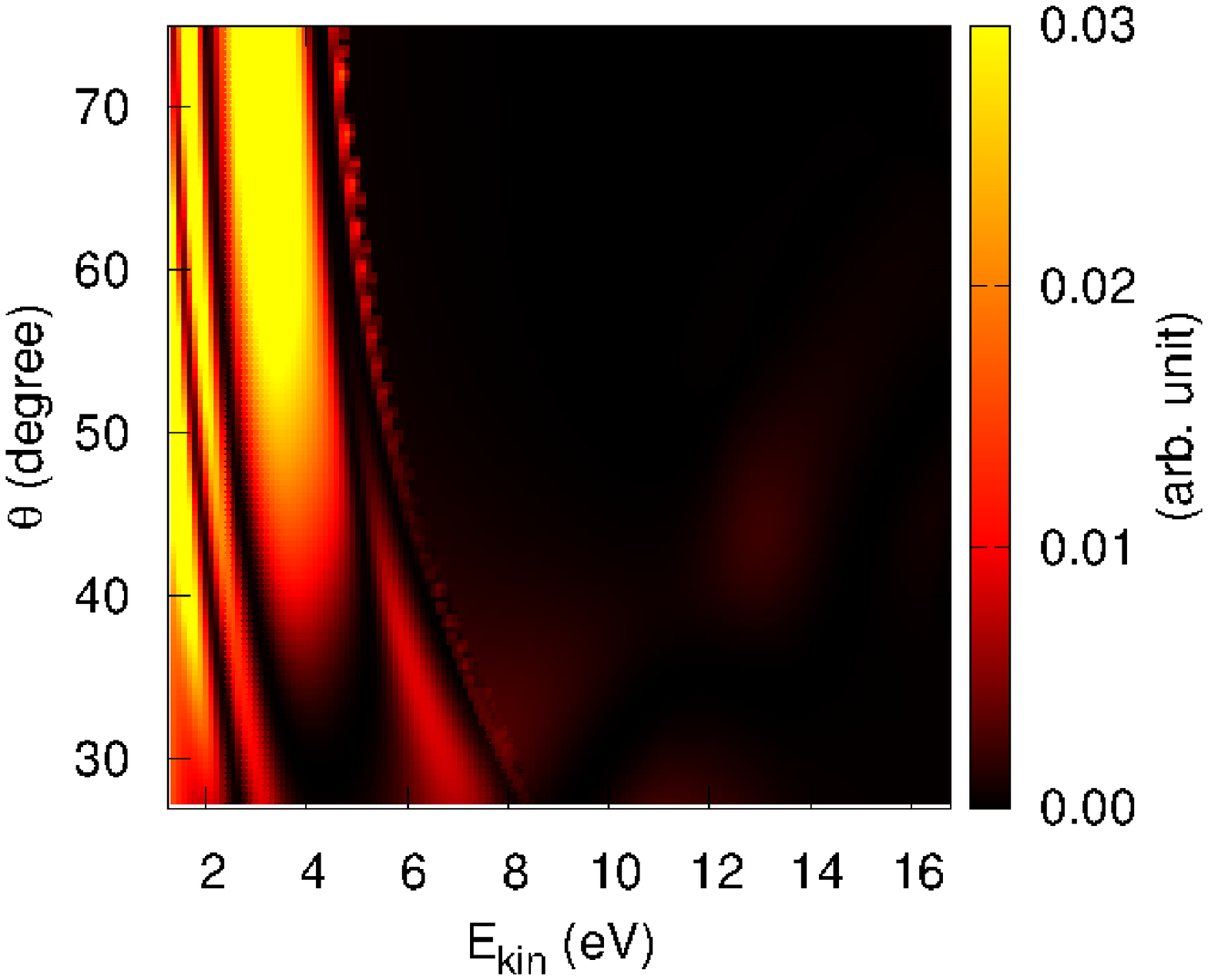}
\end{minipage}
\hspace{0.5cm}
\begin{minipage}[c]{.2\textwidth}
\includegraphics[scale=0.3,angle=270,clip=true,trim=2cm 6.2cm 1.cm 4cm]{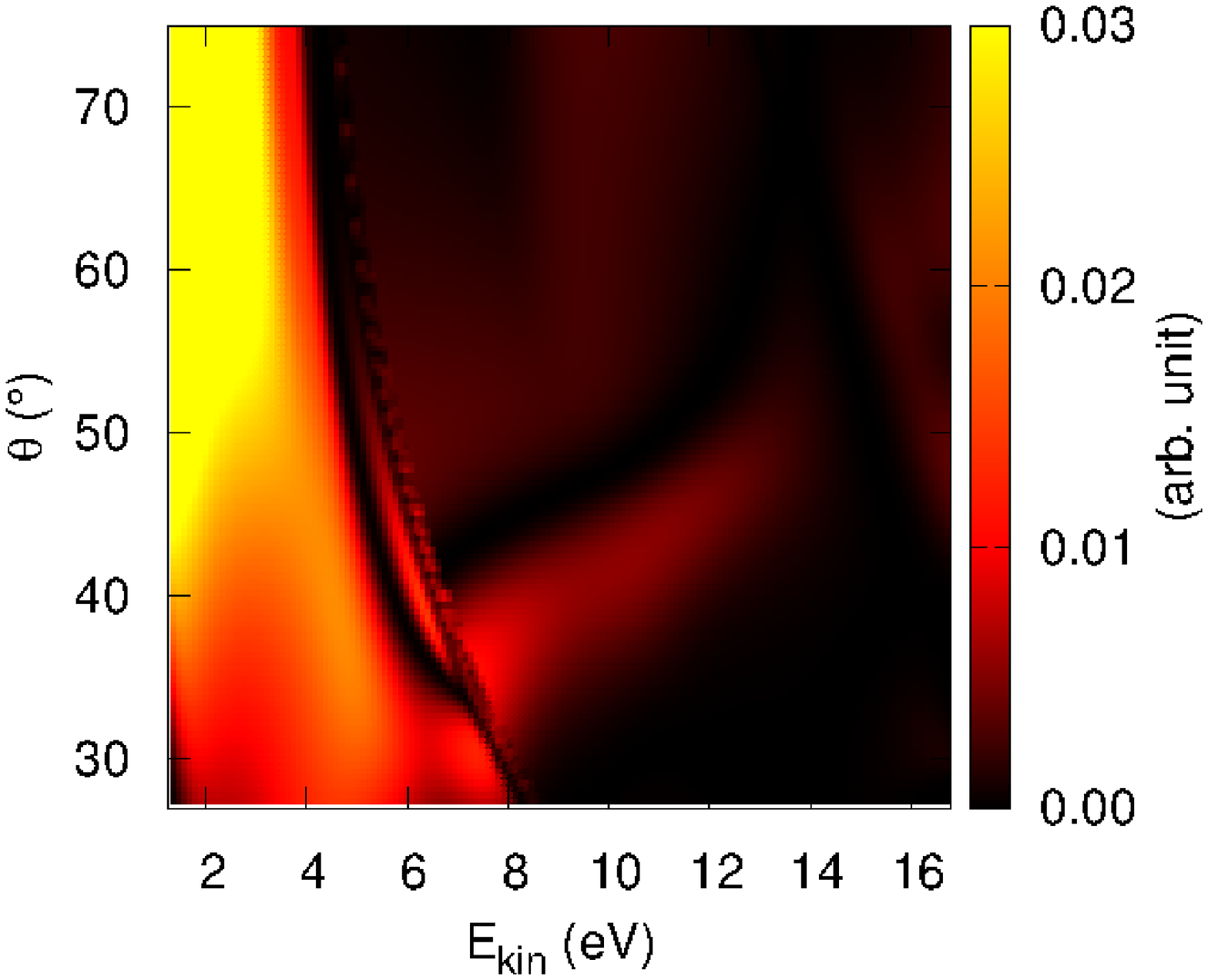}
\end{minipage}
\caption{Left: $\Theta$-energy-map of the exchange FOM for Fe(001). Right: The same for Fe(001)-p(1x1)-O.}
\label{theta_energy_maps_fom_p_m}
\end{figure}

For the SPLEED calculations, the scattering plane was aligned along the [100] direction whereas the surface magnetization
as well as the polarization of the electron was aligned along the [$\pm$100] direction.
All calculations were done for the specularly reflected beam, i.e. the (0,0) beam using the
the surface potential barrier of Rundgren-Malmstr\"om \cite{rundgren1}. For the O passivated Fe(001) surface we
calculated a work function of 7.07 eV. In comparison to a clean Fe(001) surface a value of
5 eV was calculated, i.e. an increase of the work function by passivation was found.
It should be mentioned that the work function of Fe(001) shows reasonable agreement with
experimental and theoretical values in the literature \cite{skriver1}.
We calculated polar angle ($\theta$) - energy maps that are shown in Fig.
\ref{theta_energy_maps_refl} - Fig. \ref{theta_energy_maps_fom_p_m} for 
the reflectivity, the exchange asymmetry and the FOM, respectively.
The polar angle was varied in the range from
27-75$^{\circ}$ whereas the energy range was set to 1.3 to 17 eV
according to the possible working areas as scattering mirror.

In the right panel of Fig. \ref{theta_energy_maps_refl} the effective reflectivity
of Fe(001)-p(1x1)-O is shown.
As can be seen we get a huge reflectivity especially for relatively low
kinetic energies over the full range of polar angles.
In particular at a kinetic energy of 6 eV and a polar angle
of 30$^{\circ}$ a maximum of the reflectivity occurs. Also visible in Fig.
\ref{theta_energy_maps_refl} is the emergence threshold starting around 4 eV
and a polar angle of 75$^{\circ}$ and ending at 8 eV and 27$^{\circ}$ which
marks the occurrence of a new beam.
It divides the map into mainly two parts of higher and lower reflectivity.
This results from the fact that for kinetic energies above the emergence threshold
the additional scattering channel lowers the intensity of the specular beam
as shown in recent experiments \cite{thiede1}.
The left panel in Fig. \ref{theta_energy_maps_refl} shows the effective reflectivity
for a clean Fe(001) surface. For the passivated Fe surface
higher values for kinetic energies greater than the emergence threshold have been obtained.
This is due to the higher exchange scattering for the oxygen passivated Fe surface
coming out of the higher magnetic moments at the topmost atomic layers.

In Fig. \ref{theta_energy_maps_asym_p_m_fe001_fe001_O}
the exchange asymmetry for Fe(001) and Fe(001)-p(1x1)-O, respectively, is shown.
The plots include different areas according to the preferred reflected spin-orientation 
for a defined orientation of the magnetization. 

The positive values of the exchange asymmetry correspond to a parallel
alignment of the electron spin and the sample magnetization. 
For Fe(001)-p(1x1)-O in the energy region
below the emergence threshold the scattering of parallel aligned electron spin and
magnetization is preferred except for a small area located at the emergence
threshold. For polar angles larger then 50$^{\circ}$
a crossing of the emergence threshold results in a change of the scattering behaviour.
In this case the reflected spin direction can be rotated by changing the
kinetic energy of the diffracted electrons.
For kinetic energies higher than 6 eV the scattering of electrons with antiparallel
spin alignment is preferred. Beside of a change of the polarization of the electron
a change of the magnitude occurs. The values of the calculated exchange asymmetry
fit well to the experimental data \cite{thiede1}.


Comparing the results for the asymmetry of Fe(001) and Fe(001)-p(1x1)-O
one notes that for the complete range of energy and polar angles
the asymmetry changes. This is due to the different magnetic
properties of the Fe(001) and the Fe(001)-p(1x1)-O surfaces.
At kinetic energies below the emergence
threshold the asymmetry for Fe(001)-p(1x1)-O shows broader areas with
one specific asymmetry direction. 
This is in line with to other investigations made for these systems \cite{bertacco1}.

In Fig. \ref{theta_energy_maps_fom_p_m} the FOM is shown which is the most
important observable for characterizing a material to be used as possible spin-filter.
On the left side the FOM for Fe(001) is shown whereas on the right side
for Fe(001)-p(1x1)-O. For the
oxygen passivated surface a broad range of a high FOM for kinetic
energies lower than the emergence threshold occurs. This comes from the asymmetry
$A_{+}$ (see Fig. \ref{theta_energy_maps_asym_p_m_fe001_fe001_O}) according to Eq. (\ref{eq:fom}).
It is an advantage for the application
as spin filter using low kinetic energy electron diffraction
for the determination of surface properties.
For Fe(001)-p(1x1)-O the highest values are reached for kinetic energies lower than 4 eV and a
polar angle larger than 50$^{\circ}$. Due to the fact that working
points for spin-filters are suitable between 40$^{\circ}$ and 60$^{\circ}$ the oxygen
passivated surface is a promising candidate for such an application.

\begin{figure}[h]
\hspace{-1cm}
\begin{minipage}[c]{.2\textwidth}
\includegraphics[scale=0.3,angle=270,clip=true,trim=2cm 4.2cm 1.cm 7.7cm]{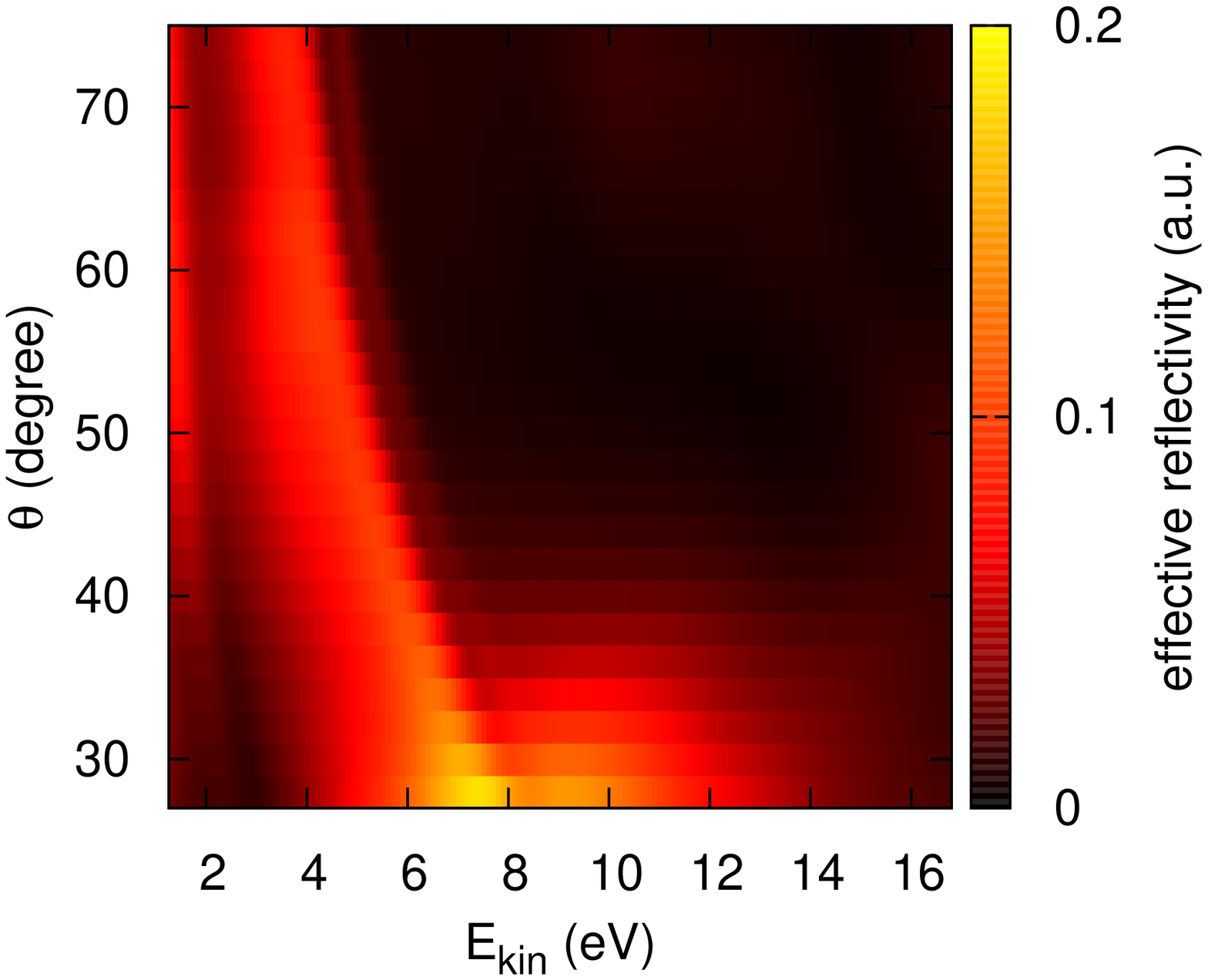}
\end{minipage}
\hspace{0.5cm}
\begin{minipage}[c]{.2\textwidth}
\includegraphics[scale=0.3,angle=270,clip=true,trim=2cm 7.0cm 1.cm 3cm]{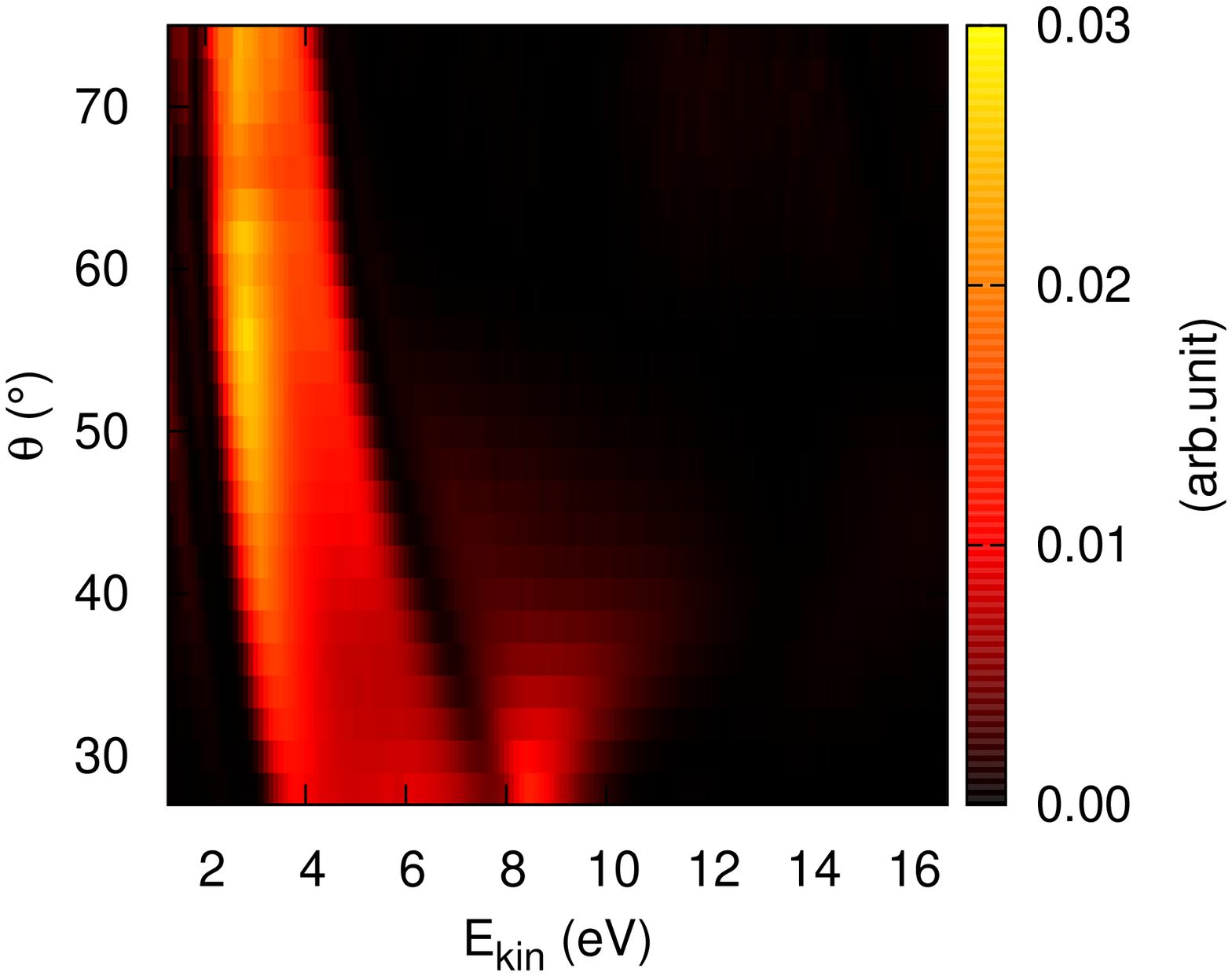}
\end{minipage}
\caption{Experimental results for the SPLEED measurements on Fe(100)-p(1x1)-O taken from Ref. \cite{thiede1}. (Reproduced by permission.)
Left: $\Theta$-energy-map of the reflectivity. 
Right: $\Theta$-energy-map of the exchange FOM.}
\label{theta_energy_maps_fom_p_refl_p_experiment}
\end{figure}

\begin{figure}[H]
\begin{minipage}[c]{.2\textwidth}
\includegraphics[scale=0.3,angle=270,clip=true,trim=2cm 4.cm 1.cm 7.7cm]{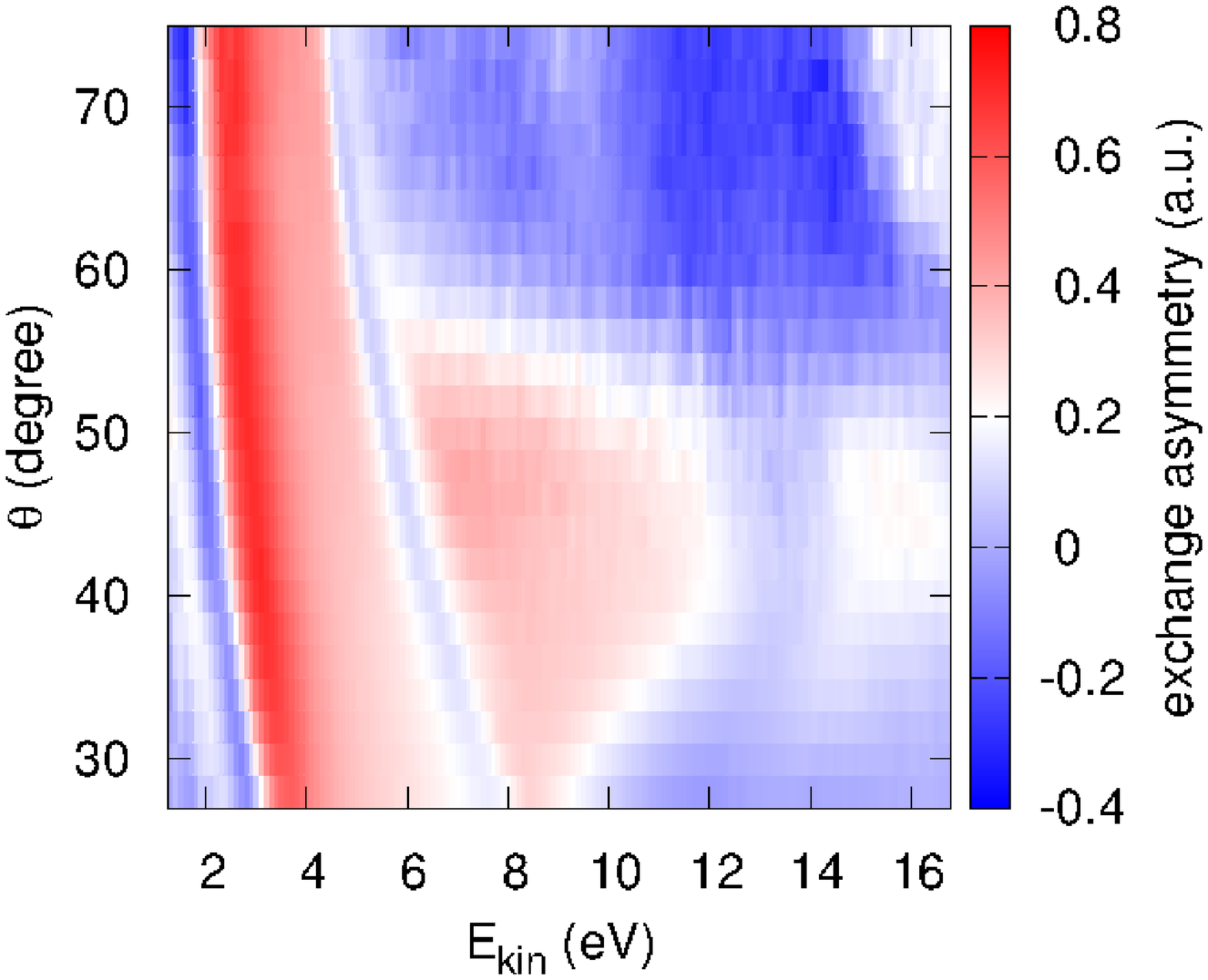}
\end{minipage}
\hspace{0.5cm}
\begin{minipage}[c]{.2\textwidth}
\includegraphics[scale=0.3,angle=270,clip=true,trim=2cm 6.2cm 1.cm 4cm]{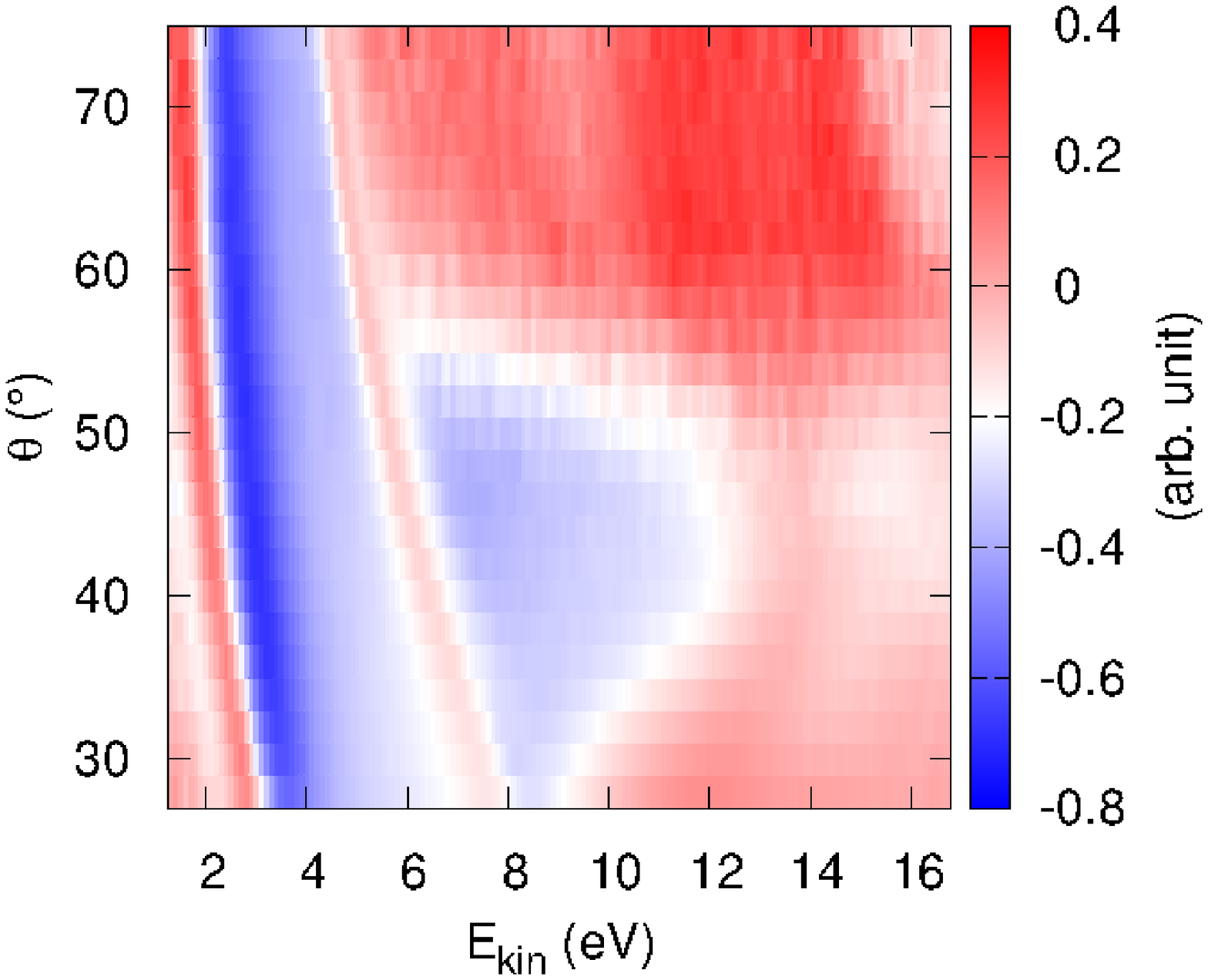}
\end{minipage}
\caption{Experimental results for the SPLEED measurements on Fe(100)-p(1x1)-O taken from Ref. \cite{thiede1}. (Reproduced by permission.)
Left: $\Theta$-energy map of the exchange asymmetry.
Right: $\Theta$-energy map of the exchange asymmetry for the reversed
magnetization.}
\label{theta_energy_maps_asym_p_m_experiment}
\end{figure}

In Figs. \ref{theta_energy_maps_fom_p_refl_p_experiment} and \ref{theta_energy_maps_asym_p_m_experiment} the
experimental results for the reflectivity, the FOM and the exchange asymmetry are shown for comparison \cite{thiede1}.

The emergence threshold is well reproduced by the theoretical results, i.e. its correspondence for varying
polar angle and kinetic energy. Hence the geometric configuration described by our
calculations match the setup in the experiment. Also the inner potential calculated out of the work function and the
Fermi energy is confirmed. This is ensured by the fact that a difference to the experiment would result in a energy shift of the
emergence threshold. In Fig. \ref{theta_energy_maps_asym_p_m_experiment}
the change of the exchange asymmetry is shown by reversing the magnetization of the Fe(100)-p(1x1)-O surface.
The same behavior as in the theoretical results are visible coming out of vanishing
spin-orbit asymmetry. Based on that changing the magnetization direction exactly
inverses the scattered polarization of the electrons.

\subsection{Many-body effects in SPLEED calculations}
\begin{figure}[H]
\begin{minipage}[c]{.2\textwidth}
\includegraphics[scale=0.3,angle=270,clip=true,trim=2cm 4.cm 1.cm 7.5cm]{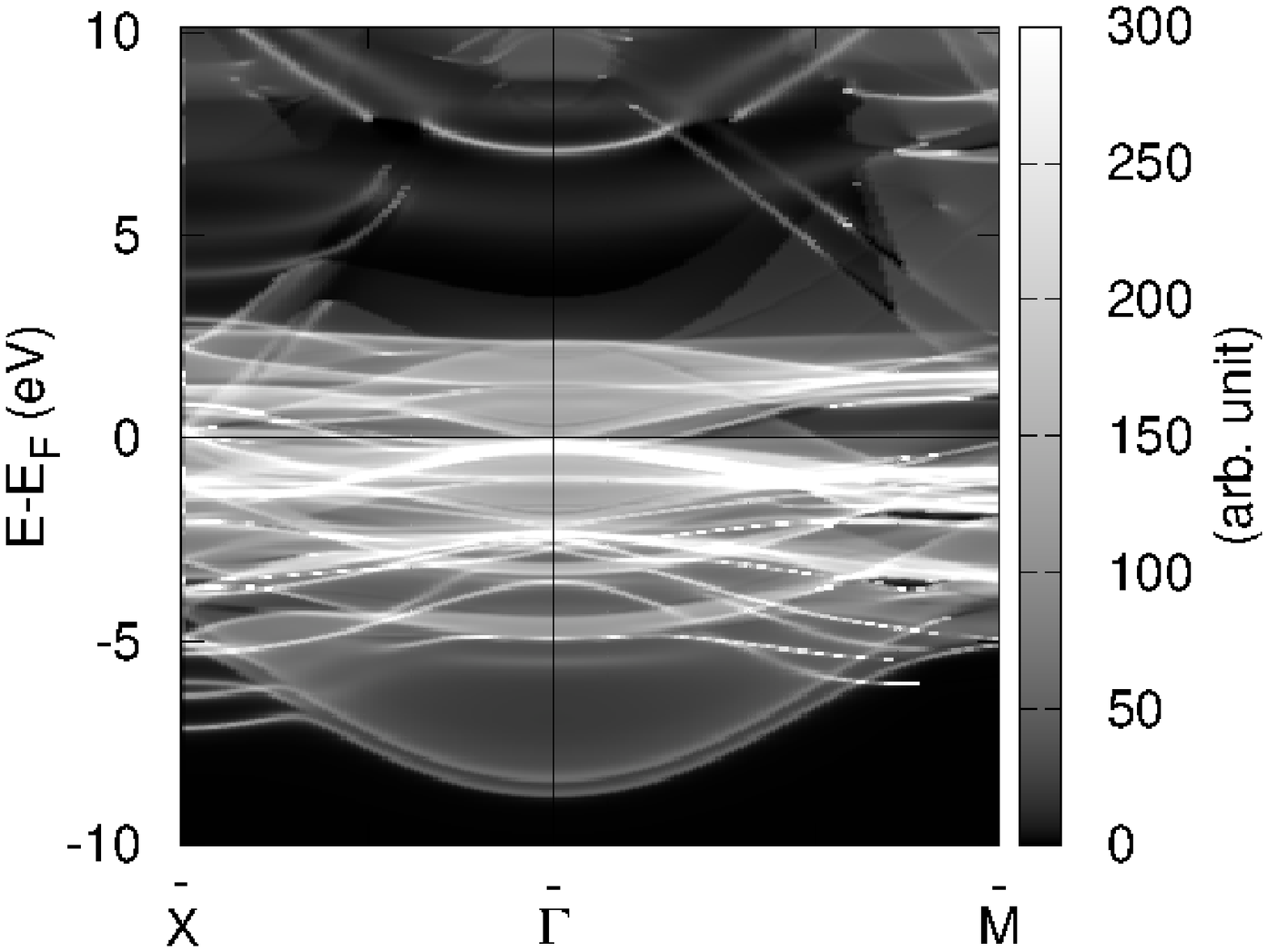}
\end{minipage}
\hspace{0.5cm}
\begin{minipage}[c]{.2\textwidth}
\includegraphics[scale=0.3,angle=270,clip=true,trim=2cm 6.2cm 1.cm 7.5cm]{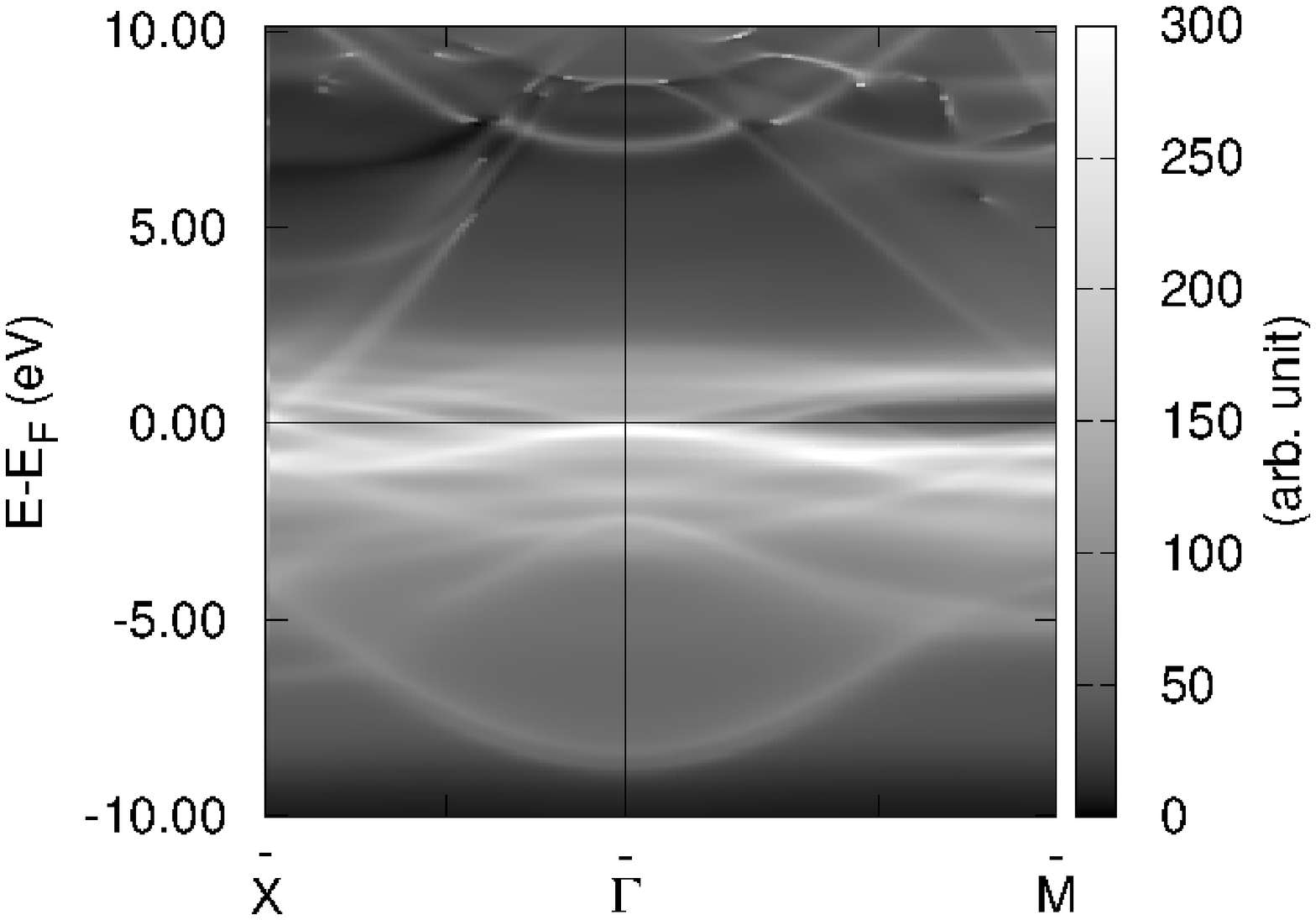}
\end{minipage}
\caption{Left: Bloch spectral function without consideration of many body effects
         for Fe(001)-p(1x1)-O.
         Right: Results including many body effects via the DMFT for Fe(001)-p(1x1)-O.}
\label{bsf_dmft}
\end{figure}

\begin{figure}[H]
\centering
\includegraphics[scale=0.3,angle=270,trim=1cm 0cm 1cm 0cm,clip=true]{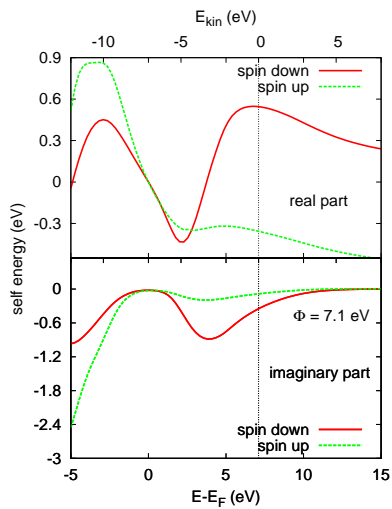}
\caption{Real and imaginary part of the self energy for both spin channels.}
\label{selfenergy_ImSig}
\end{figure}

The importance of the inclusion of many body effects for spectroscopic calculations
has been shown in different works \cite{braun1,minar1,haule1,marco1,gray1}. 
Because of the low kinetic energy of the incident electrons many-body effects
might become important for the spectroscopic calculations due to changes
of relevant bands (surface states, bands at the Fermi energy, unoccupied states)
resulting from a change of the underlying electronic structure calculations.
In Fig. \ref{bsf_dmft} the band structure 
of Fe(001)-p(1x1)-O with and without inclusion of the DMFT is shown. 
As one notes, the bands around 7 eV above the Fermi level
are smeared out by many body interactions.
This energy range is important for
SPLEED giving characteristic spectral features.
Changes in the band dispersion results in different
magnetic properties altering the exchange interaction at the sample surface.
This effect can be connected to the self energy
shown in Fig. \ref{selfenergy_ImSig}.
Comparing to the calculated work function
(7.07 eV) the self energy has a nonzero value affecting the valence bands relevant
for the exchange scattering process. Although our calculations show that the impact
on the effective reflectivity and the exchange asymmetry is negligible, the changes in
the FOM are significant.

\begin{figure}[H]
\begin{minipage}[c]{.2\textwidth}
\includegraphics[scale=0.3,angle=270,clip=true,trim=2cm 4.cm 1.cm 7.5cm]{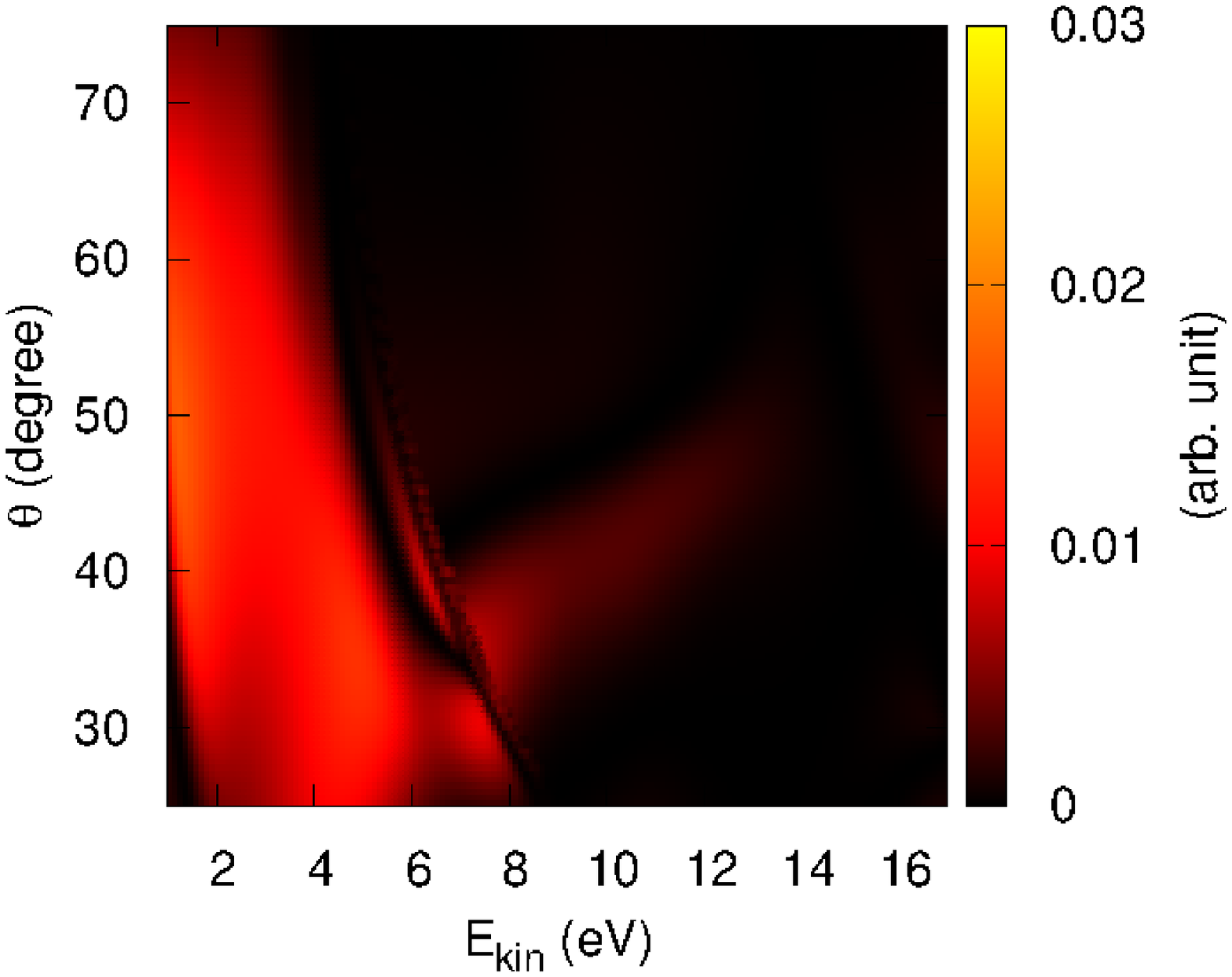}
\end{minipage}
\hspace{0.5cm}
\begin{minipage}[c]{.2\textwidth}
\includegraphics[scale=0.3,angle=270,clip=true,trim=2cm 7.2cm 1.cm 4cm]{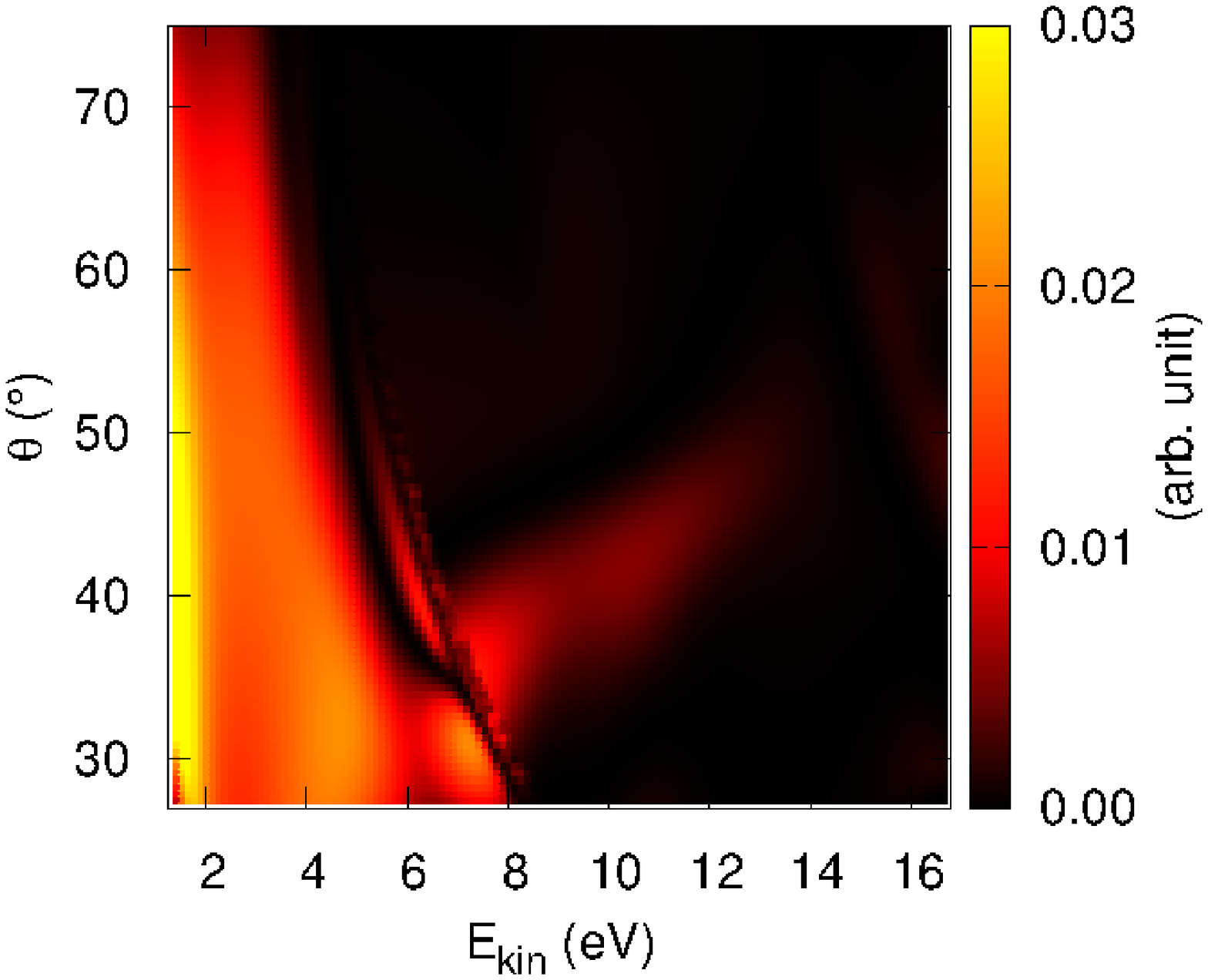}
\end{minipage}
\caption{FOM without (left) and with (right) consideration of many body effects
         for Fe(001)-p(1x1)-O.
         In both cases the projection of the electron spin has been included.}
\label{theta_energy_maps_fom_p_m_dmft}
\end{figure}

In Fig. \ref{theta_energy_maps_fom_p_m_dmft} we present the FOM for Fe(001)-p(1x1)-O
resulting from LSDA+DMFT-based SPLEED-calculations.
Additionally we considered the change in the projection of the polarization 
of the electron concerning the surface when changing the polar angle.
In comparison to Fig. \ref{theta_energy_maps_fom_p_m} this results
in a shift of the maximal value of the FOM to a polar angle around $50^\circ$.
According to the changes of the band structure which affects essentially the
band near the Fermi level, changes in the FOM are seen mainly for low kinetic
energies. This is important comparing measurements and calculations of 
spectra for low energy electron diffraction.

\section{Summary \label{sec_summary}}
We have shown that the calculations done using our ab initio method
regarding the SPLEED-spectra for Fe(001)-p(1x1)-O are in
satisfying agreement with recent experimental results \cite{thiede1}.
Therefore our description of the systems electronic properties seem
to be confirmed. The system exhibits a large FOM
and various suitable areas for the application as spin-polarizing mirror.
We have shown that a projection of the polarization of the electron has
a huge impact on the exchange scattering, especially for
the calculation of the FOM.
Furthermore the inclusion of many body effects
has been considered for the first time in SPLEED calculations
showing that this results in spectral changes important for
the regime of very low energy electron scattering.

\section{Acknowledgement}
We thank the BMBF (05K13WMA), the DFG (FOR 1346),
CENTEM PLUS (LO1402) and the COST Action MP 1306 for financial support.
We also thank the group of Prof. M. Donath at the Westf\"alische Wilhelms-Universit\"at
M\"unster for fruitful discussions and for the permission
to include the experimental results (all taken from Ref. \cite{thiede1}).

\bibliographystyle{/home/stephan/revtex/revtex4-1/bibtex/bst/revtex/apsrev4-1}

\bibliography{spleed_fe001_O}

\end{document}